\documentclass[reprint,authaffil,NumberedRefs,12pt]{JASAnew}

\usepackage{setspace}

\usepackage{enumitem}
\usepackage{amsmath}
\usepackage{cases}
\usepackage{xcolor}
\usepackage[linesnumbered,ruled,vlined]{algorithm2e}
\usepackage{tikz}

\SetCommentSty{mycommfont}
\begin{document}

\title[Multiscale CNN for DML]{Multiscale CNN based Deep Metric Learning for Bioacoustic Classification: Overcoming Training Data Scarcity Using Dynamic Triplet Loss}
\author{Anshul Thakur}
    \author{Daksh Thapar}
    \author{Padmanabhan Rajan}
    \author{Aditya Nigam}
\affiliation{School of Computing and Electrical Engineering, IIT Mandi, Mandi, Himachal Pradesh-175005, India}
\email{anshul\_thakur@students.iitmandi.ac.in}
\preprint{Anshul Thakur, JASA}		

\date{\today} 

\begin{abstract}
This paper proposes multiscale convolutional neural network (CNN)-based deep metric learning for bioacoustic classification, under low training data conditions. The proposed CNN is characterized by the utilization of four different filter sizes at each level to analyze input feature maps. This multiscale nature helps in describing different bioacoustic events effectively: smaller filters help in learning the finer details of bioacoustic events, whereas, larger filters help in analyzing a larger context leading to global details. A dynamic triplet loss is employed in the proposed CNN architecture to learn a transformation from the input space to the embedding space, where classification is performed. The triplet loss helps in learning this transformation by analyzing three examples, referred to as triplets, at a time where intra-class distance is minimized while maximizing the inter-class separation by a dynamically increasing margin. The number of possible triplets increases cubically with the dataset size, making triplet loss more suitable than the softmax cross-entropy loss in low training data conditions. Experiments on three different publicly available datasets show that the proposed framework performs better than existing bioacoustic classification frameworks. Experimental results also confirm the superiority of the triplet loss over the cross-entropy loss in low training data conditions. 

\end{abstract}


\maketitle

\section{\label{sec:1} Introduction}

Habitat destruction induced by global warming and human activities has pushed many avian and amphibian species to the brink of extinction. With this looming threat of population decline and species extinction, a huge escalation can be witnessed in the conservation efforts \cite{birds,cushman2006effects}. The surveying and monitoring are the principal steps in any conservation effort. Manual surveying and monitoring are difficult, time-consuming and require experienced personnel \cite{borker2014, furnas2015using}. Owing to the rich acoustic communication in birds and frogs, automated acoustic monitoring provides an appropriate way to survey different species of interest in their natural habitat and alleviates the requirement of manual monitoring \cite{brandes2008automated}. The bioacoustic signal classification module is the mainstay of such acoustic monitoring systems \cite{ssa_mlsp} and often includes tasks such as bird and frog species classification. The major impediment in many bioacoustic classification tasks is the scarcity of the labeled training data. Moreover, the target species and hence, the training data requirements often vary from one ecosystem to other. This makes it unfeasible to collect and label a large amount of bioacoustic data for all the possible species. Thus, there is a requirement for bioacoustic classification frameworks that could provide effective classification in low training data conditions.

In recent times, deep convolution neural networks (CNN) have become the cornerstone for achieving state-of-art performances in various audio classification tasks \cite{cakir_taslp,xu2018large,hershey2017cnn}. In comparison to the shallow learning techniques, CNNs often require large amount of training data (subject to the task in hand) to generalize and provide effective classification. However, the scarcity of the labeled data for many bioacoustic tasks makes it undesirable to utilize these data-intensive CNNs. The lesser amount of training data often leads to over-fitting in CNNs. This over-fitting can be avoided by using regularizers and early stopping, which can restrict the modeling capabilities of the CNNs. Many studies on CNN based audio classification have used data augmentation techniques to overcome the training data scarcity \cite{salamon2017deep,lu}. These methods augment the training data with synthetic examples that is generated by deforming the original data. Some common deformations used for data augmentation include pitch alterations and time stretching. However, these augmentation techniques are not always useful and can affect the classification performances \cite{lu}. As a result, the effectiveness of these techniques are data dependent and often require a trial-error approach. In the case of bioacoustics, coming up the effective augmentation requires domain-expert knowledge about the nature of vocalizations of each target species. Apart from augmentation, many studies have explored transfer learning for overcoming training data scarcity \cite{dielmann,ntalampiras2018bird}. In the case of CNNs, existing (or pre-trained) networks trained for any audio classification tasks can be fine-tuned for achieving effective performance \cite{finetuning}. Fine-tuning helps in transferring the knowledge from the pre-trained network to the domain and task of interest \cite{poria2016convolutional}. In low training data conditions, fine-tuning an existing network is easier and effective than training the network from scratch. Thus, CNN based transfer learning presents an effective way to overcome the training data scarcity in bioacoustic applications.



In literature, CNNs have mildly been explored for different bioacoustic classification tasks. Due to the recently conducted \emph{bird activity detection (BAD) challenges} \cite{bad1}, large datasets have been publicly released for the task of bird activity detection. This led to an influx of CNN-based frameworks that provide state-of-art performance for the aforementioned task \cite{cakir,grill,densenet}. However, due to the scarcity of the labeled data, only a few studies have addressed the task of vocalization segmentation and species identification using deep learning approaches. Lostanlen \emph{et al.}\cite{lostanlen2018birdvox}~ released a bird flight call detection dataset along with a CNN-based benchmark. Salamon \emph{et al.}\cite{salamon2017fusing} experimentally showed that the late fusion of scores obtained from CNN (deep learning) and a random forest classifier (shallow learning) results in better performance for the task of bird species classification from flight calls. The same CNN architecture, consisting of three convolution layers and two dense layers, is used in both the aforementioned studies. Ibrahim and Zhuang\cite{ibrahim2018automatic} proposed to use a recurrent neural network (RNN) and CNN to classify grouper species. T{\'o}th\cite{toth}, Sprengel \emph{et al.}\cite{sprengel}~and Piczak\cite{piczak} utilized spectrogram enhancement methods before applying CNNs to identify bird species from their songs or calls. This spectrogram enhancement helps in removing the effect of overwhelming background disturbances on the classification procedure. 

Apart from deep learning, many classical machine learning techniques have been successfully utilized for bioacoustic classification. Stowell and Plumbley\cite{dan_skmeans} proposed spherical K-means based unsupervised feature learning for large scale bird species classification. Building on their work, Thakur \emph{et al.}\cite{CCSE,dcr} proposed to use archetypal analysis\cite{CVPR_AA} and deep archetypal analysis for obtaining supervised convex representations for bioacoustic classification. Kernel-based extreme learning machines are used by Qian \emph{et el.}\cite{eml}~ for bird species classification. This study utilizes active learning to alleviate the problem of unlabelled bioacoustic data. Many studies have used dynamic kernels (such as the intermediate matching kernel and probabilistic sequence kernels) based support vector machines (SVM) for different bioacoustic classification tasks such as bird activity detection and bird species classification\cite{deep,rpsk,aa_BAD}.         

In this work, the authors propose to use CNN-based deep metric learning (DML)\cite{yi2014deep} for bioacoustic classification. DML deals with learning a mapping from the input space to a compact Euclidean space where similarity among examples is in direct correspondence with distance among them. As a result, DML directly provides class-specific clustering in this space. Thus, a classifier trained in this space can provide better classification than the one trained in the input feature space. This study utilizes CNN powered by the triplet loss\cite{facenet} to map the input examples to the 128-dimensional embeddings in the desired transformation space. The triplet loss processes three examples, called a triplet, at a time. Triplet consists of an anchor, a positive example and a negative example. The anchor and the positive examples belong to the same class whereas the negative example can be from any other class. A CNN with triplet loss tries to learn a transformation where a triplet constraint is imposed on all the training examples. This constraint states that the distance between the negative-anchor pair should be greater than the positive-anchor by a fixed margin in the transformation space. Only triplets that violate this constraint are chosen for training. More details about the triplet loss and its implementation are in later sections. Although triplet loss has been successfully utilized for many applications such as face recognition\cite{facenet} and person re-identification\cite{yi2014deep}, it has also received some criticism for its slow convergence\cite{magnet} on large datasets. This slow convergence can be attributed to the fact that as the size of dataset increases, the possible number of triplets increase cubically\cite{pahariya2018dynamic}. This drawback of triplet loss can prove advantageous in the low training data conditions as the possible number of triplets even in small datasets can be large enough to effectively train a CNN. Moreover, the nature of each triplet (in terms of distance between positive-anchor and negative-anchor pairs) is usually unique. As a result, no redundant information is utilized in the training procedure.

CNN used in the proposed DML framework is characterized by the utilization of different filter sizes in the convolution layers. Each filter size helps in analyzing the input bioacoustic events at a different scale. The smaller filters help in extracting the minute local details whereas the large filters analyze a larger receptive field and help in obtaining the global details from the input bioacoustic event. This notion of multiscalar analysis is inspired by the Inception\cite{incept} model that was proposed for large scale image classification. This multiscale CNN is empowered by a dynamic variant of the classical triplet loss\cite{facenet} to learn the desired transformation space. During training, the margin of the loss function is slowly increased based on a pre-defined heuristic (see section 2). This dynamically varying triplet loss has a two-fold advantage: 
\begin{enumerate}
    \item Starting with a smaller margin and slowly increasing the margin can be seen as the warm start. First, the CNN is taught to learn a relatively simpler task of separating the examples of one class from the others in the embedding space by a smaller margin. Then, the complexity of this task is slowly increased by increasing the margin. This warm start helps in better convergence even when the number of classes is very large.
    \item Dynamically varying margin increases the number of triplets used for training. It can be attributed to the fact that triplets which satisfy the triplet constraint at a lower margin can violate the constraint as the margin is increased.
\end{enumerate}
 
The main contribution of this study are as follows:

\begin{itemize}
\item To the best of authors' knowledge, this is the first study that utilizes deep metric learning for bioacoustics.
\item A simple multiscale CNN is proposed for bioacoustic classification.
\item This study experimentally shows that the utilization of triplet loss helps in overcoming the training data scarcity without utilizing any data augmentation and transfer learning. 
\item A dynamic variant of triplet loss is proposed. 
\end{itemize}

The rest of this paper is organized as follows.  In Section \ref{sec:prop}, the proposed DML based classification framework and the dynamic triplet loss is described. Experimental setup is explained in Section \ref{sec:setup}. Experimental results are in discussed in Section \ref{sec:results}. Section \ref{sec:con} concludes this paper.

\begin{figure}[t]
	\centering
	\includegraphics[trim={1cm 1cm 1cm 0cm},scale=0.45]{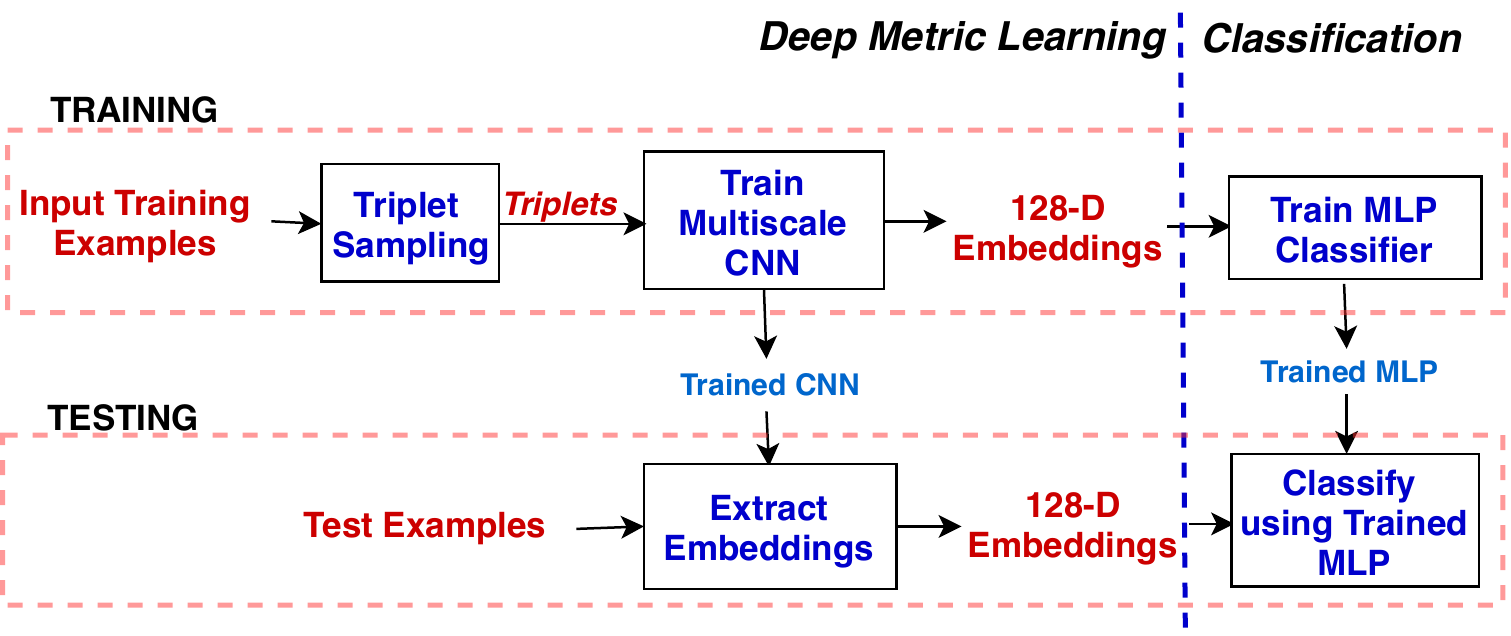}
	\caption{Proposed DML framework for bioacoustic classification.}
	\label{fig:dml_class}
\end{figure}

\section{Proposed DML Framework}
\label{sec:prop}
In this section, the proposed DML framework for bioacoustic classification is described. The overall design of the framework is depicted in Fig.~\ref{fig:dml_class}. The proposed framework is composed of two neural networks: a multiscale CNN and a multi-layer perceptron (MLP). The multiscale CNN, equipped with dynamic triplet loss, is used to learn a transformation from input to the embedding space. The embeddings generated by CNN are given as input to the MLP for learning the discrimination between classes in the embedding space. 

This section starts with the feature extraction procedure. Then, the architectures of the proposed multiscale CNN and MLP are described. Later, dynamic triplet loss and other details regarding the training of neural networks are highlighted. Finally, the procedure to classify the bioacoustic signals using the trained DML framework is explained.

\begin{figure*}[t]
	\centering
	\includegraphics[trim={1cm 0.5cm 0cm 0cm},scale=0.5]{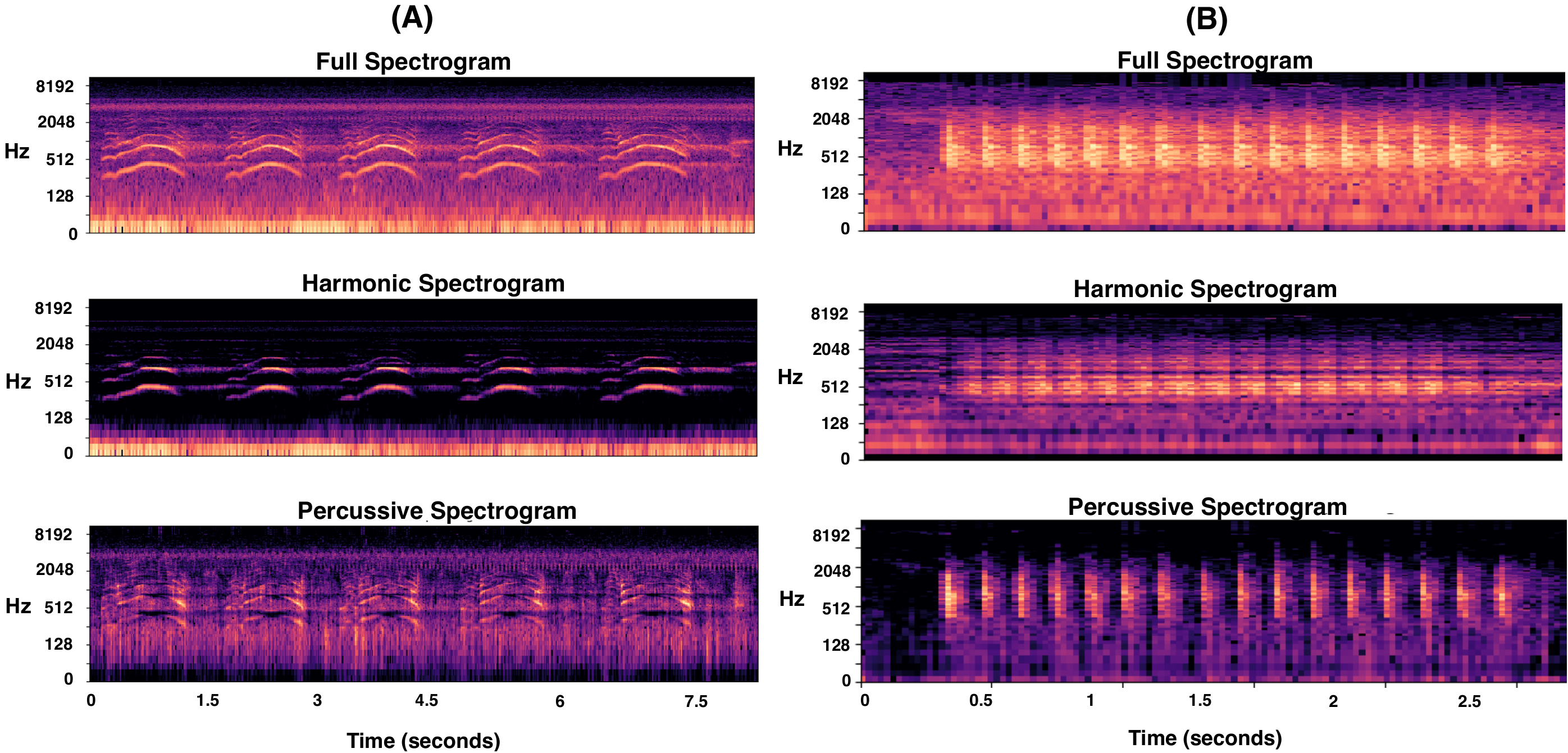}
	\caption{Difference in harmonic and percussive components of sounds produced by (A) Indian peafowl and (B) white-bellied woodpecker.}
	\label{fig:fig1}
\end{figure*}

\subsection{Feature Extraction}
Most bird species such as Passerines are known for producing the harmonically rich sounds. However, there are many species such as woodpeckers, snipes and storks that are characterized by drumming, winnowing, clattering and other mechanically produced sounds. These sounds are more or less percussive in nature. Thus, the difference in harmonic and percussive components of a bioacoustic sound has some class-specific characteristics. This difference in harmonic and percussive components of sounds produced by white-bellied woodpecker and Indian peafowl is evident in Fig.~\ref{fig:fig1}. Inspired by this observation, Mel-spectrogram along with its harmonic and percussive components\cite{driedger2014extending} are given as a three-channel input to the proposed framework. The spectrogram of the input audio recording is decomposed into its harmonic and percussive components using the method proposed in Ref. \hyperlink{target}{39}. The original, harmonic and percussive spectrograms are multiplied by Mel filterbank to obtain the respective Mel spectrograms that form the three channels of an input example. All three channels are converted to decibel scale and are normalized with respect to the maximum value.

\begin{figure*}[t]
	\centering
	\includegraphics[trim={4cm 0.75cm 0cm 1cm},scale=0.47]{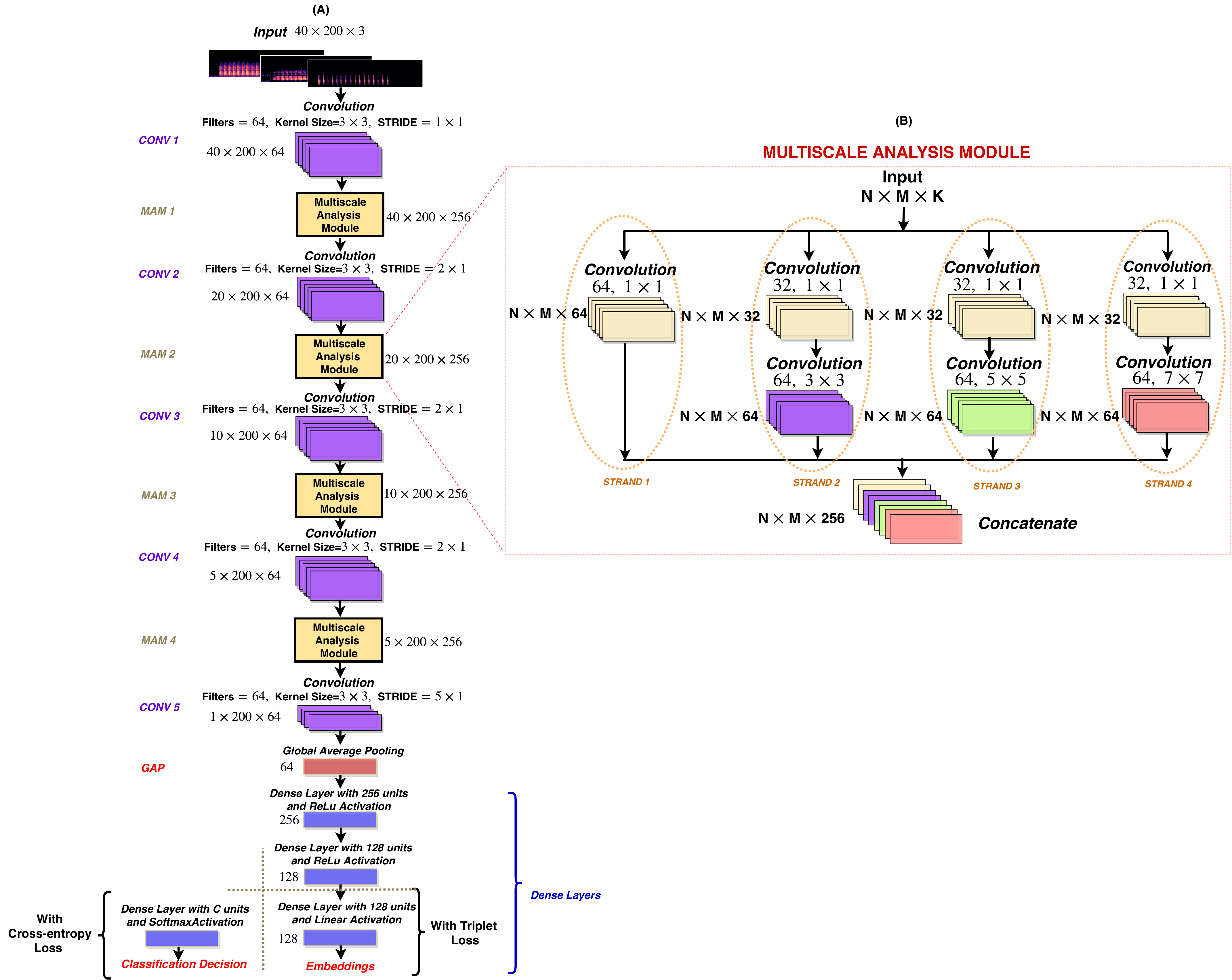}
	\caption{Illustration of (A) the Proposed multiscale CNN architecture and (B) a multiscale scale analysis module. To use the proposed multiscale CNN architecture for classification (with cross-entropy loss), the last layer of the architecture is replaced by a fully connected layer having $C$ (number of classes) units and softmax activation. 
	}
	\label{fig:fig2}
\end{figure*}

\subsection{Neural Network Designs}
\subsubsection{Multiscale CNN Architecture}
The proposed CNN consists of five convolution (\emph{CONV}) layers, four Inception\cite{incept} inspired multiscale analysis modules ($\emph{MAM})$, three dense layers and has 1,286,410 trainable parameters. Each multiscale analysis module consists of seven convolution layers having different filter sizes that enable the network to analyze each input at different scales. The overall network design is illustrated in Fig.~\ref{fig:fig2}(a). The main components of the network are:

\noindent\textbf{\emph{Input:}} As discussed earlier, audio examples, represented by Mel-spectrograms and their harmonic and percussive components ($40\times M\times3$, 40 Mel-filters and $M$ frames) are given as input.

\noindent\textbf{\emph{Multiscale Analysis Module:}} 
Multiscale analysis modules utilize kernels of different sizes ($1\times1$, $3\times3$. $5\times5$ and $7\times7$). This multiscale analysis helps in better feature extraction from short duration vocalizations (such as birdsong syllables or flight calls) as well as from the longer vocalizations such as birdsong phrases. The shorter vocalizations occupy smaller spatial space on Mel-spectrograms as compared to the longer vocalizations. Hence, a smaller kernel size is more appropriate for the shorter bioacoustic events and vice-versa. The smaller $3\times3$ filter helps in learning minute details from an input Mel-spectrogram where as the larger filters ($5\times5$ and $7\times7$) help in capturing more global traits due to the larger receptive fields. In bioacoustics, these minute details can be low energy harmonics or vocalizations recorded in a far-field setting. The global traits can include the information about the frequency contents or bandwidth and the coarser time-frequency modulations of bioacoustic events.

\begin{figure*}[t]
	\centering
	\includegraphics[trim={3cm 2.75cm 3cm 1cm},scale=0.55]{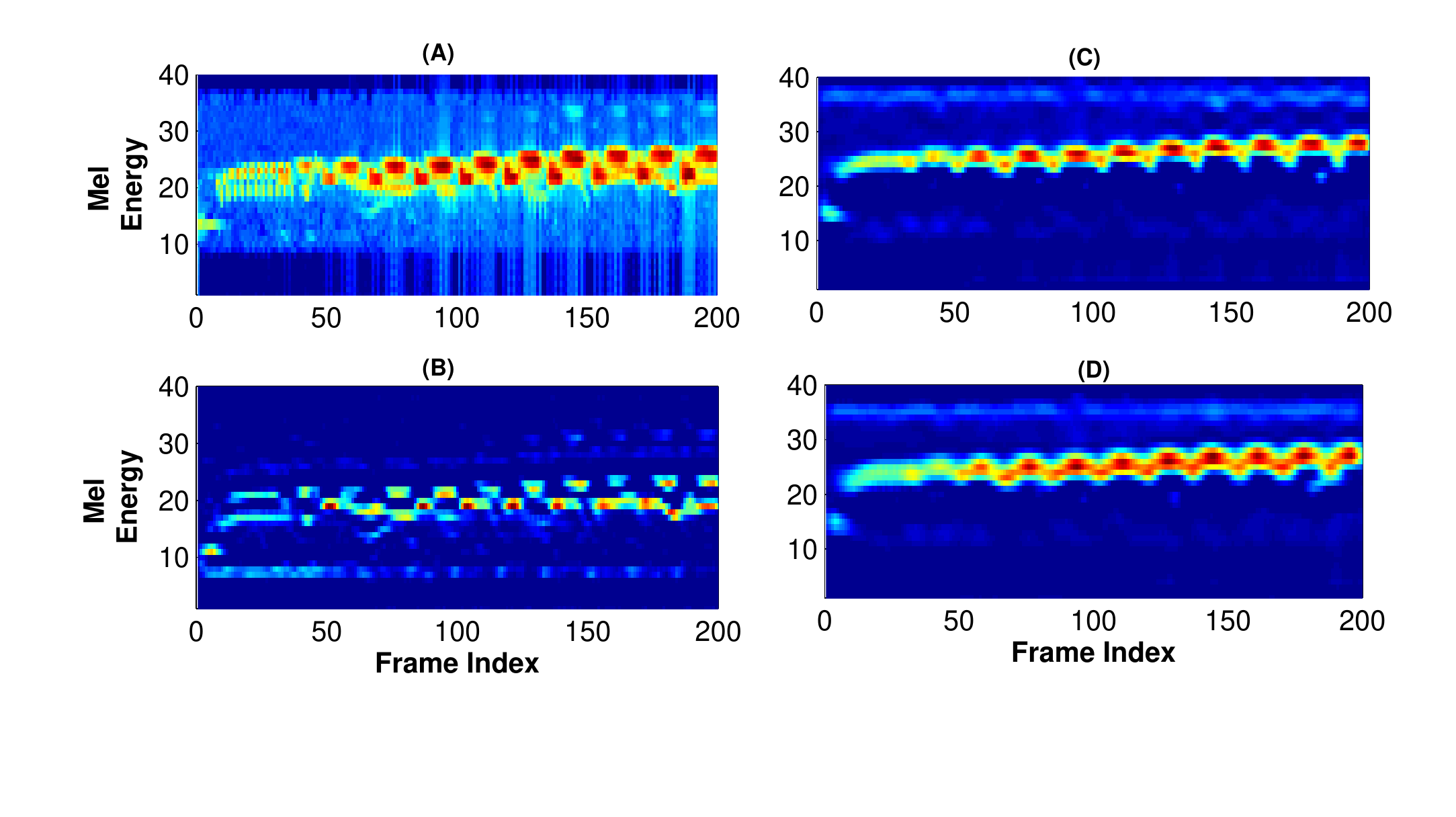}
	\caption{(A) Mel-spectrogram given as input to the proposed CNN. (B), (C) and (D) depicts the filter responses obtained for $6^{th}\; 3\times3$, $32^{nd}\; 5\times5$ and $62^{nd}\; 7\times7$ filters of the first multiscale analysis module (\emph{MAM 1}) of the trained multiscale CNN. These particular filters are chosen for their expressivity.   
	}
	\label{fig:fig3}
\end{figure*}

Each \emph{MAM} receives an input of 64 feature maps ($N\times M\times64$) that are processed by seven convolution layers arranged in four parallel strands as shown in Fig.~\ref{fig:fig2}(b). First strand contains one convolution layer that has 64 filters of $1\times1$ kernel size. These filters are mainly concerned with selecting the inter feature map patterns rather than the spatial analysis of feature maps. It is a known fact that each feature map has some complementary information\cite{all_conv}. Hence, learning these
inter feature map patterns can be helpful in distinguishing one class from another. The second, third and fourth strands consist of two convolution layers. The first convolution layer in all these strand consists of 32 filters of $1\times1$, whereas, the second convolution layers have 64 filters of sizes $3\times3$, $5\times5$ and $7\times7$ respectively. Here the $1\times1$ convolution layers serve two purposes: 1). It decrease the number of input channels from 64 to 32 and reduce the computational requirements for the following convolutional layer in each strand. 2). As discussed earlier, $1\times1$ filters are used for selecting the discriminative feature patterns from the input feature maps. It must be noted that each strand utilizes a separate $1\times1$ convolution layer. As the appropriate feature patterns may be scale dependent, a separate $1\times1$ convolution layer provides independence in the feature selection for different scales in each parallel strand. The output of these layers are processed by convolution layers having filter size of $3\times3$, $5\times5$ and $7\times7$ in second, third and fourth strand respectively for multiscale analysis. The difference in responses of filters of different strands is illustrated in Fig.~\ref{fig:fig3}. The feature maps obtained from all the four strands are concatenated in a channel-wise manner to output 256 feature maps from each module. It must be noted that zero-padding is used to make sure that each feature map is of same dimension before concatenation.            

\noindent\textbf{\emph{Bottleneck, Global Pooling and Dense Layers:}} The network consists of five convolution layers having 64 filters of $3\times3$. Apart from the first convolution layer, all other convolution layers act as the bottleneck layers. They use strided convolutions to down-sample the feature maps by a factor of 2 (or 5 in case of the last convolution layer) along the Mel-energy axis. Apart from down-sampling, they also help in selecting the relevant features from the feature maps, obtained from multiscalar analysis modules, by analyzing the inter feature map correlations\cite{all_conv}. Due to this feature selection, the number of channels are decreased from 256 (generated by multiscale analysis module) to 64 which also helps in decreasing the computation requirements for the corresponding layers. After all the convolution layers and multiscale analysis modules, global average pooling (GAP) is applied to obtain a 64-D vector. This averaging operation helps in making the framework invariant towards the time differences in onsets-offsets of the bioacoustic events in audio recordings or their Mel-spectrograms. Then, this 64-D vector is passed to the dense layers. The network has three dense layers having 256, 128 and 128 hidden units.     

\noindent\textbf{\emph{Activation, Regularization and Optimizer:}} Each convolution layer (whether stand-alone or in multiscale analysis module) and first two dense layers are followed by rectified linear unit activation. The output of the last dense layer is normalized to have the unit norm such that embeddings produced by the proposed CNN lies on a unit hypersphere. A dropout of 0.5 is added before each dense layer. Along with dropout, exponential weight decay of $0.0001$ is also used to avoid over-fitting and improve generalization\cite{krogh1992simple}. Adagrad with a learning rate $0.001$ is used as optimizer.    
 
\subsubsection{MLP Architecture}
\label{sssec:mlp}
The MLP used in the proposed framework consists of three layers: an input layer with 128 units, a hidden layer with 256 units and an output layer with $C$ units ($C$ is the number of classes). The rectified linear unit function is used as activation for the hidden layer neurons. The weight optimization is performed by utilizing Adam solver. A constant learning rate of 0.001 and L2 regularization term of 0.0001 is used for optimization.

\subsection{Multiscale CNN Training: Dynamic Triplet Loss}
Dynamic triplet loss is utilized for training the proposed multiscale CNN.  
The aim of the triplet loss is to learn an embedding space where all possible triplets satisfy the triplet constraint. Let $(\mathbf{x}^a_i,\mathbf{x}^p_i,\mathbf{x}^n_i)$ be $i$th triplet of embeddings belonging to the possible set of triplets present in the training data embeddings. $\mathbf{x}^a_i$, $\mathbf{x}^p_i$ and $\mathbf{x}^n_i$ be an anchor, a positive and a negative example. Then, the triplet constraint can be defined as: 

\begin{equation}
    \Vert\mathbf{x}^a_i-\mathbf{x}^n_i\Vert_2^2- \Vert\mathbf{x}^a_i-\mathbf{x}^p_i\Vert_2^2 \geq \alpha.
    \label{eq:1}
\end{equation}

Here $\alpha$ is the enforced margin or distance between the positive examples and the negative examples. Thus, based on the triplet constraint, the loss function to be minimized is\cite{facenet}:     
\begin{equation}
   \mathcal{L}=\sum_{i=1}^{N}\textrm{max}( \Vert\mathbf{x}_i^a-\mathbf{x}_i^p\Vert_2^2- \Vert\mathbf{x}^a_i-\mathbf{x}^n_i\Vert_2^2 + \alpha,0),
\end{equation}
where $N$ is the possible number of triplets to be used for training.

For training, triplets are sampled from a mini-batch in an online fashion. A forward pass is performed on the CNN to obtain embeddings for a mini-batch of input examples. These embeddings are processed to select triplets that are later used for optimizing the current state of CNN. The performance of triplet loss is directly dependent on the choice of triplets to be used for training. Choosing triplets that satisfy triplet constraint (equation \ref{eq:1}) will lead to no change in the state of CNN. Hence, triplets violating the triplet constraint are of interest. These triplets are of two types: hard triplets and semi-hard triplets. A triplet $\mathcal{X}=(\mathbf{x}^a,\mathbf{x}^p,\mathbf{x}^n)$ is classified as hard or semi-hard according to the following criteria\cite{facenet}:

$\mathcal{X} \;\;\textrm{is}\; \begin{cases}
   hard & d(\mathbf{x}^a,\mathbf{x}^n)<d(\mathbf{x}^a,\mathbf{x}^p) 
   \\
   semi-hard & d(\mathbf{x}^a,\mathbf{x}^p)<d(\mathbf{x}^a,\mathbf{x}^n)\; \textrm{and} 
   \\ & d(\mathbf{x}^a,\mathbf{x}^n)<d(\mathbf{x}^a,\mathbf{x}^p)+\alpha.
  
\end{cases}$
\vspace{0.5cm}

Here $d()$ refers to the Euclidean distance.

Although hard triplets appear to be more informative for training, they result in higher loss values, leading to the larger weight updates. These larger weight updates result in significant change to the current state of network, hence, undoing the optimization work done by the previous weight updates. Thus, utilizing these hard triplets may lead to instability during training. It has been shown in Ref. \hyperlink{tar}{35} that semi-hard triplets often leads to faster convergence and effective training than the hard triplets. Building on this information, the semi-hard triplets are used for training the proposed CNN. In semi-hard triplet, the distance between anchor-negative pairs is greater than the anchor-positive pairs as desired. However, this distance is not greater than the desired separation margin $\alpha$. Thus, the weight updates obtained in case of semi-hard triplets are not as large as the hard triplets, leading to a stable training.

As discussed in Section \ref{sec:1}, a dynamic variant of the triplet loss is used in this work. In the proposed implementation of triplet loss, $\alpha$ or the margin is considered as a dynamic variable whose value is changed over the course of training. The overall procedure to calculate dynamic triplet loss is depicted in Algorithm \ref{1}.

\begin{algorithm}[h]
\SetAlgoLined
\label{1}
 \caption{Training CNN using Dynamic Triplet Loss}
 \SetKwData{Left}{left}
\SetKwData{This}{this}
\SetKwData{Up}{up}
\SetKwFunction{Union}{Union}
\SetKwFunction{FindCompress}{FindCompress}
\SetKwInOut{Input}{input}
\SetKwInOut{Output}{output}
\Input{$f()$: CNN (randomly initialized)
$X$: Training dataset}
\vspace{0.1cm}
\Output{$f()$: Trained CNN for metric learning}
\BlankLine

 $\alpha=0.2$ \tcp{Initial value of margin}
 $count\_list=[\;]$ \tcp{List to store number of triplets sampled in each iteration}
 $thresh=15$ \tcp{Threshold for margin update (determined empirically)}
 $K=100$ \tcp{Number of epochs}
  \For{$J\gets1$ \KwTo $K$}{
   $\mathcal{I}=\textrm{createBatches}(X,n)$ \tcp{Returns n batches stored in List $\mathcal{I}$ }
      \For{$i\gets1$ \KwTo $n$}{
        $\mathcal{E}=f(\mathcal{I}[i])$ \tcp{Forward pass to get embeddings for $i$th batch}
         $\mathcal{T},t=\textrm{getTriplets}(\mathcal{E},\alpha)$
         \tcp{Returns $\mathcal{T}$, a set containing $t$ semi-hard triplets, sampled from $i$th batch}
         $count\_list.\textrm{append}(t)$ \tcp{Store the number of semi-hard triplets sampled from $i$th batch}
         $num=len(count\_list)$ \tcp{Number of elements in $count\_list$ currently}
         \If{$num\geq3$ AND $\alpha\leq0.6$}
         {
         \If{($count\_list[num]<thresh$ \textrm{AND} $count\_list[num-1]<thresh$ \textrm{AND} $count\_list[num-2]<thresh$)}
         {$\alpha=\alpha+0.05$ \tcp{Margin update}}
         }

         $L=\textrm{calculateTripletLoss}(f(),\mathcal{T},\alpha)$ \tcp{Calculate Triplet Loss using Equation 2}
        
         $f()=\textrm{UpdateWeights}(f(),L)$ \tcp{Back-propagate $L$ through $f()$ to get the updated $f()$
        }
         }
  }
\end{algorithm}

We start with a small margin, $\alpha=0.2$, and force the network to learn the embedding space where examples of each class are separated from other by a distance of $0.2$. As the network is trained, the number of semi-hard triplets mined from training dataset decreases. If this number of mined triplets is less than a pre-defined threshold for three consecutive iterations, the value of $\alpha$ is incremented by $0.05$. Again, the network is trained to satisfy the new triplet loss induced by new value of $\alpha$. This process is continued till $\alpha$ reaches a pre-defined maximum value of $0.6$. As discussed in Section \ref{sec:1}, this dynamic triplet loss can lead to faster convergence. Along with threshold, values of $\alpha$ and margin updates are determined experimentally.

   \begin{figure*}[t]
	\centering
	\includegraphics[trim={2cm 0cm 2cm 0cm},scale=0.5]{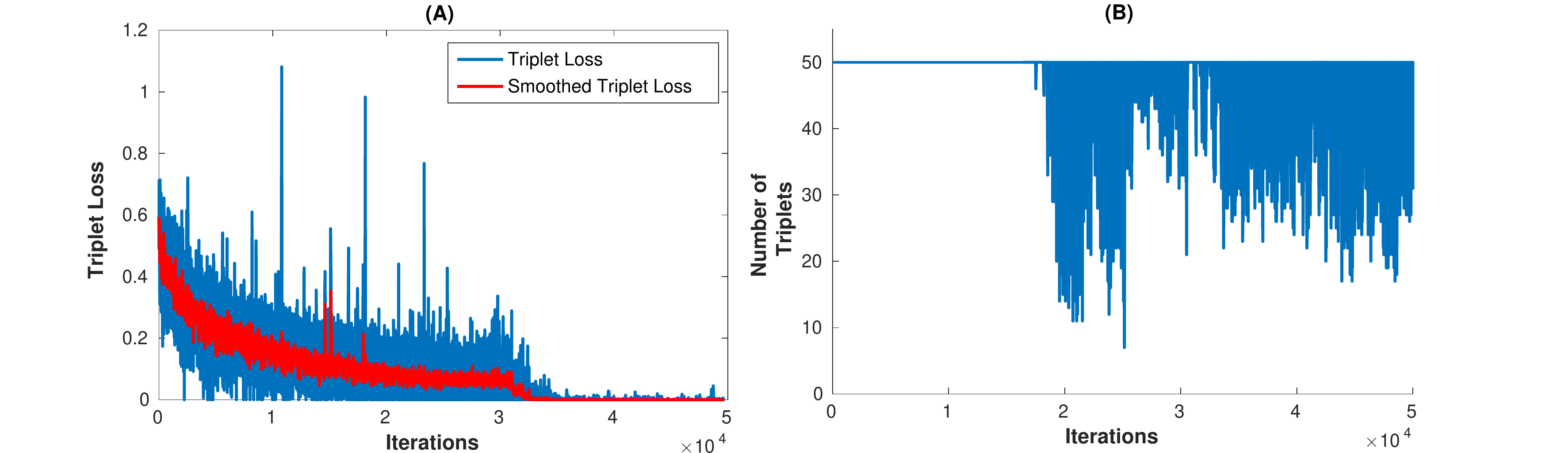}
	\caption{Illustration of (A) triplet loss and (B) number of semi-hard triplets involved during during each iteration of training the multiscale CNN with CLO-43SD dataset. The red plot in (A) represents the smoothed triplet loss over a window of 15. 
	}
	\label{fig:triplet_stats}
\end{figure*}

\emph{\textbf{Implementation Details}}:
Each mini-batch is composed of at least $5$ examples per class. Hence, all classes are represented in a mini-batch. The semi-hard triplets are sampled from this mini-batch and are used for training the CNN. The number of triplets that can be processed simultaneously (let us say triplet batch size) is often limited by the available GPU memory. In our implementation, we set this triplet batch size to be 150 input examples or 50 semi-hard triplets. Thus, the semi-hard triplets sampled from a mini-batch are presented to the CNN in an iterative manner where during each iteration, the triplet batch (having 50 or less triplets) is used to calculate triplet loss and update the weights. The number of mini-batches in an epoch are limited to 1000. The CNN is trained till it reaches the desired value of margin ($\alpha$) and the triplet loss is converged to zero. Figure \ref{fig:triplet_stats} depicts the number of triplets and triplet loss calculated from each triplet batch during the training of multiscale CNN on CLO-43SD dataset (see Section \ref{sec:setup} for dataset details).

\subsection{Classification}
\begin{figure}[t]
	\centering
	\includegraphics[trim={2cm 0.75cm 2cm 0cm},scale=0.5]{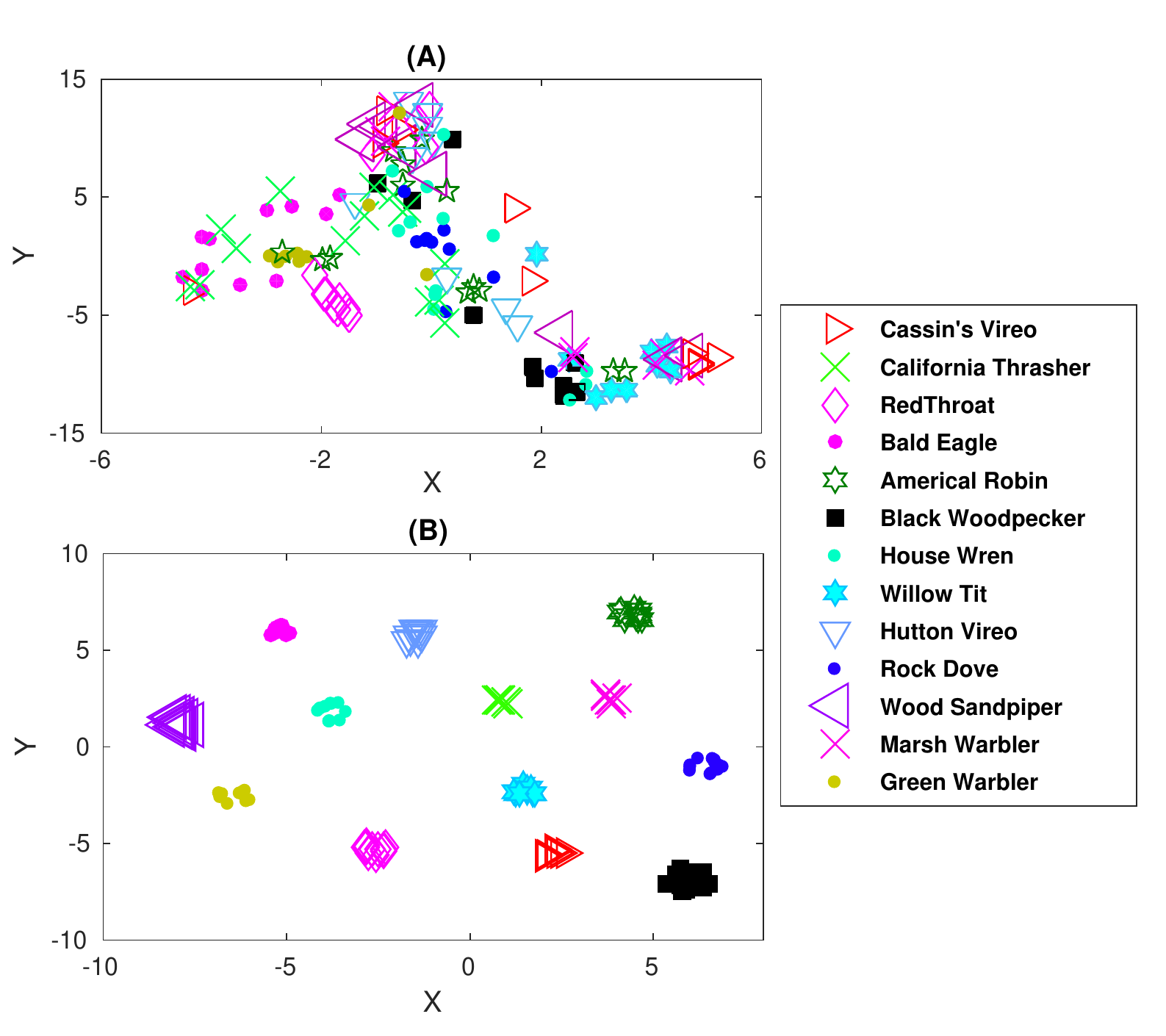}
	\caption{Two dimensional t-SNE visualization of 128-d embeddings extracted from audio examples of 12 different bird species using (A) untrained multiscale CNN and (B) fully trained multiscale CNN. 
	}
	\label{fig:tsne}
\end{figure}

As illustrated in Fig.~\ref{fig:dml_class}, first the multiscale CNN is trained to learn the transformation or embedding space using dynamic triplet loss. This trained CNN is used to extract 128-D embeddings from all the training examples. These embeddings show high class-specific signatures as embeddings of each class occupy a compact Euclidean space which is separated from every other classes by a significant margin. This behaviour is highlighted in Fig.~\ref{fig:tsne}. Then, a multi-layer perceptron (MLP) with Adam optimizer (described in Section \ref{sssec:mlp}) is trained on these embeddings for classification. During testing phase, embeddings are extracted from the test examples using the trained CNN. These embeddings are classified using the trained MLP classifier.       

\section{Experimental Setup}
\label{sec:setup}
In this section, we describe datasets, comparative methods, and parameter setting for the performance evaluation. 

\subsection{Datasets Used}
The performance of the proposed DML framework is evaluated on three different datasets:

\begin{itemize}[leftmargin=*]
    \item \textbf{Birdcalls71}: This dataset consists of audio recordings of 71 different bird species that are obtained from three different sources. The recordings of $38$ bird species were provided by the Macaulay Library\footnotemark[1]\footnotetext{\url{http://www.macaulaylibrary.org}} on an academic license. The recordings of $7$ bird species were downloaded from bird database maintained by Art \& Science Centre, UCLA\footnotemark[2]\footnotetext{\url{http://artsci.ucla.edu/birds/database.html}}. The recordings of $26$ bird species were obtained from the Great Himalayan national park (GHNP) dataset\cite{deep} and were provided on request. All these recordings are sampled at 44.1 kHz and vary in duration from 0.5 to 320 seconds. Due to licensing issues, the authors can not make this dataset public. However, the processed Mel-spectrograms extracted from these recordings are hosted on a public platform for analysis\footnotemark[3]\footnotetext{\url{https://figshare.com/s/4af71d71d94e04afcd5f}}.
    
    \item \textbf{Anuran dataset}: The publicly available Anuran dataset contains audio recordings of 10 different frog species found in the Amazon rainforest. These recordings are sampled at 44.1 kHz and are of variable durations (from 3 to 360 seconds). More details about the frog species and dataset can in found in Ref.~\hyperlink{frog}{6}.   
   
     \item \textbf{CLO-43SD}: This public dataset is provided by Salamon \emph{et al.}\cite{salamon2016towards} and contains processed Mel-spectrograms of flight calls of 43 different North American wood-warblers. These Mel-spectrograms were extracted from audio recordings sampled at 22.05 kHz using 11.6 ms frame size with an overlap of 1.25 ms and 40 Mel bands. It must be noted that 11.6 ms frame size is optimum to analyze flight calls\cite{salamon2016towards,salamon2017fusing}.  
     
     \item \textbf{Combined}: To analyze the scalability of the proposed framework, all three datasets are combined together to create a larger dataset having 124 classes.

\end{itemize}

\begin{table}[h]
\caption{Comparative methods used for the performance evaluation.}
\label{comp_meth}
\resizebox{0.45\textwidth}{!}{ 
\begin{tabular}{|c|c|}
\hline
\textbf{Method}                                                                                  & \textbf{Nature}                                                                  \\ \hline
\begin{tabular}[c]{@{}c@{}}Spherical K-means and\\  Random forest (SKM)\cite{dan_skmeans}\end{tabular}             & \begin{tabular}[c]{@{}c@{}}Unsupervised \\ Feature Learning\end{tabular}         \\ \hline
\begin{tabular}[c]{@{}c@{}}Deep Convex  \\ and Random Forest (DCR)\cite{dcr}\end{tabular}                  & \begin{tabular}[c]{@{}c@{}}Supervised \\ Dictionary Learning\end{tabular}        \\ \hline
\begin{tabular}[c]{@{}c@{}}Kernel Based Extreme \\ Learning Machines (KELM)\cite{eml}\end{tabular}         & \begin{tabular}[c]{@{}c@{}}Shallow Learning\end{tabular} \\ \hline
VGG\cite{hershey2017cnn}                                                                                              & CNN                                                                \\ \hline
\begin{tabular}[c]{@{}c@{}}Fine-tuned VGG (VGG-FT)\cite{hershey2017cnn},\\ Pre-trained on AudioSet\end{tabular}      & \begin{tabular}[c]{@{}c@{}}Transfer Learning/CNN\\ /Deep Learning\end{tabular}   \\ \hline
\begin{tabular}[c]{@{}c@{}}CNN proposed by \\ Salamon et al.\cite{salamon2017deep} (SAL)\end{tabular}                  & CNN                                                                \\ \hline
\begin{tabular}[c]{@{}c@{}}Multiscale CNN with Cross \\ Entropy Loss  (MS-CNN)\end{tabular}      & CNN/Deep Learning                                                                \\ \hline
\begin{tabular}[c]{@{}c@{}}Multiscale CNN with Triplet Loss \\ and MLP  (MS-CNN-TL)\end{tabular} & \begin{tabular}[c]{@{}c@{}}CNN based Deep \\ Metric Learning\end{tabular}        \\ \hline
\end{tabular}}
\end{table}

\subsection{Comparative studies}
The classification performance of the proposed DML framework is compared with six existing bioacoustic and audio classification frameworks. Table \ref{comp_meth} lists different methods used for the performance comparison. These comparative methods include both shallow and deep learning based methods. The three mentioned shallow learning baselines include polynomial kernel based extreme learning machines (KELM)\cite{eml} and random forest classifiers, wherein (1) the KELM is trained on low-level audio descriptors while the later one is trained on (2) the unsupervised and (3) supervised feature representations. The unsupervised feature representations are obtained from spherical K-means\cite{dan_skmeans} (SKM) while the supervised representations\cite{dcr} are acquired by deep convex matrix factorization (DCR). The input feature representations (Mel-spectrogram and compressed spectral frames) used in the respective studies are also used here. In SKM, frame-wise feature representations for each input example are aggregated using mean and standard deviation to obtain a fixed dimensional representation. In DCR, a random forest classifier is trained on a frame-wise deep convex representation and a voting rule is used on these frame-wise decisions to classify the input example. 

Apart from these shallow techniques, CNN proposed by Salamon \emph{et al.}\cite{salamon2017fusing} (SAL) and VGG (used in Ref.~\hyperlink{vgg}{9} for audio classification) are used as deep learning baselines. The total number of trainable parameters in SAL and VGG are 1,226,554 and 9,653,831 respectively (if last dense layer has 71 units). To evaluate the proposed DML framework against transfer learning, we fine-tuned VGG network for bioacoustic classification. This VGG network is pre-trainied on AudioSet database (see Ref.~\hyperlink{vgg}{9} for more details). The dense layers of VGG are changed with three dense layers having 256, 128 and $C$ (number of classes) hidden units. A dropout of 0.5 is used before each dense layer. The first two dense layers have Relu activation where as the last dense layer is followed by softmax activation. The final baseline is the proposed multiscale CNN with the cross-entropy loss (MS-CNN). The last dense layer of the propose CNN is replaced with a dense layer having $C$ units and softmax activation ($C$ is the number of target classes). The Keras implementations of these CNN baselines are publicly available along with datasets\footnotemark[3].

\subsection{Parameter Setting}
All the audio recordings used for experimentation are divided into fixed length segments of 2 seconds. These segments are used for training and performance comparison. If the duration of any recording is less than 2 seconds, then the signal is repeated (from the beginning) to force the fixed duration of 2 seconds. The short-term spectral analysis is done using a 20 ms frame size with 50\% overlap to obtain respective feature representations (from all databases except CLO-43SD) used in all the considered studies. Thus, each input example consists of 200 frames. In case of CLO-43SD dataset, frames of a pre-computed Mel-spectrogram are repeated to obtain a fixed number of frames i.e. 200 per example (for maintaining uniformity between datasets). It must be noted that there is a difference in frame sizes used for the short-term analysis done for CLO-43SD and other datasets to extract Mel-spectrograms. However, this difference is ignored to create a combined dataset for the sake of analyzing the scalability of the proposed DML framework.     

For implementing SKM (for all datasets), spherical K-means with 128 clusters and random forest with 100 trees is used. For implementing DCR, a three-level archetypal analysis based matrix factorization (with an order of 128, 64 and 32) is used to learn the class-specific dictionaries. A random forest with 100 trees is used for classification. For implementing polynomial kernel and extreme learning machine in KELM, the parameter values used in Qian et al\cite{eml} were found to be optimal for all the datasets used in this study. All the aforementioned parameters are empirically determined on a validation dataset.                  

For VGG, VGG with fine tuning (VGG-FT) and MS-CNN, a batch size of 32 and Adam optimizer with a learning rate of 0.001 is used. These model are trained for 200 epochs and checkpoints are used after each epoch to determine the setting that provides least validation loss. For implementing the network proposed in Salamon \emph{et al.} (SAL), the details and parameter setting described in the respective study is also used here. The implementation details of the proposed DML framework (MS-CNN-TL) are already discussed in Section \ref{sec:prop}. The multiscale CNN is trained using triplet loss for 150 epochs (each epoch consists of 1000 iterations). After 150 epochs, the MS-CNN-TL shows full convergence. In the current context, the convergence simply means that all semi-hard triplets in the training dataset satisfy the triplet constraint across all dynamically chosen margins.


\subsection{Train-Test Data distribution and Performance Metric}
Birdcalls71, CLO-43SD and Anuran datasets have 1982, 5428 and 2445 examples respectively.  All datasets considered in this study exhibit significant imbalance in the number of examples per class. In Birdcalls71, the number of examples per class vary from 9 to 75. Similarly, in CLO-43SD, this number varies from 10 to 1256. Thus, a single class of CLO-43SD forms about $23\%$ of the dataset. To avoid this huge class imbalance, this class is not included in the study. The number of examples per class varies from 50 to 705 in Anuran dataset. The $50\%$, $15\%$ and $35\%$ of the examples per class (in all datasets) are randomly chosen for training, validation and testing. All these files are publicly available at \emph{Figshare}\footnotemark[3]. 

\noindent\emph{Performance Metric:}  The imbalance between classes is large, however, we desire to give equal weightage to the classification performance obtained for each class. Hence, macro F1-score \cite{van2013macro} is used as a metric for the performance comparison. The macro F1-score is the average of class-specific F1-scores where F1-score is the harmonic mean of precision and recall.

\section{Results and Discussion}
\label{sec:results}
In this section, the classification performances of the proposed DML framework and different baselines are presented. Apart from that, the significance of using harmonic-percussive components and generalization of the proposed framework are also discussed. Finally, the extension of the proposed framework for open-set classification is described.

\begin{figure*}[t]
	\centering
	\includegraphics[trim={0cm 0cm 0cm 0cm},scale=0.6]{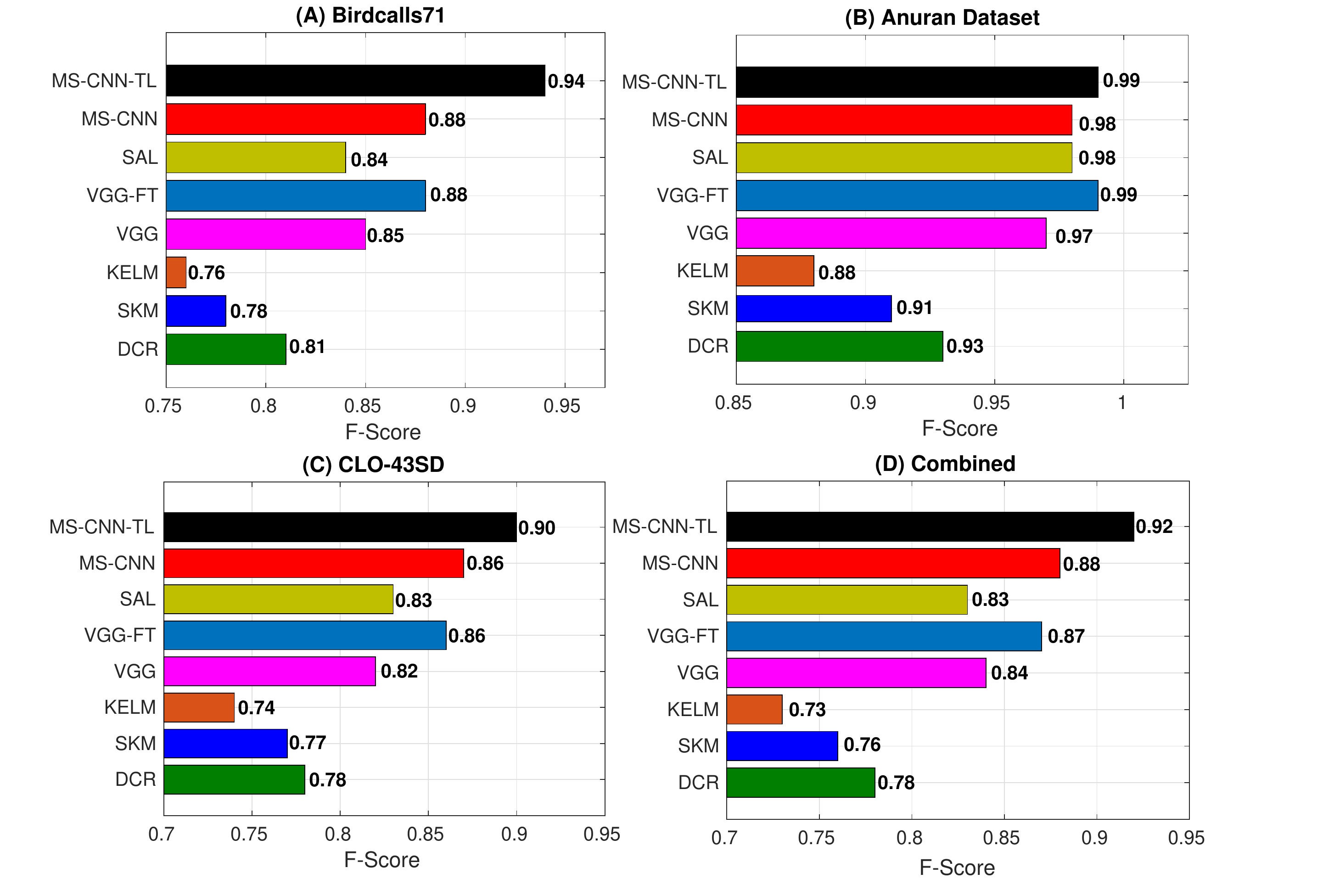}
	\caption{Classification performances of the proposed framework along with various baselines on (A) Birdcalls71, (B) Anuran, (C) CLO-43SD and (D) Combined datasets.   
	}
	\label{fig:results}
\end{figure*}
 
\subsection{Classification Performance}
Fig.~\ref{fig:results} shows the classification performances of different methods on all four datasets. Following can be inferred from the analysis of this figure:

\begin{itemize}[leftmargin=*]
    \item Shallow learning techniques (DCR, SKM and KELM) are significantly outperformed by CNN based frameworks including the proposed MS-CNN and MS-CNN-TL.
    
    \item MS-CNN performs better than VGG and SAL on all but Anuran dataset, highlighting the superiority of the proposed multiscale CNN. On Anuran dataset, all these baselines show comparable performances.
    
    \item VGG-FT outperforms VGG and SAL while showing similar performance to that of MS-CNN. This shows that the utilization of transfer learning or fine-tuning a pre-trained model improves the classification performance.    
    
    \item MS-CNN-TL (proposed) outperforms all baselines including MS-CNN on all datasets. This confirms the claim that utilizing triplet loss based DML framework leads to a better classification than the cross-entropy loss based CNN.
    Moreover, the better performance of MS-CNN-TL than VGG-FT shows that the proposed framework provides effective classification without using any fine-tuning or pre-trained model. 
    
    \item The class unbalance has no significant effect on the performance of MS-CNN-TL as evident from the high macro F1-score across all datasets. Since the target of triplet loss based metric learning is to separate examples of one class from other by a fixed margin, the difference in number of examples per class may not have a significant effect the learning procedure. Moreover, as discussed earlier, during mini-batch creation, at least $5$ examples per class are always present in each batch and each class is represented in the training procedure.  
    
\end{itemize}
\begin{figure}[t]
	\centering
	\includegraphics[trim={1.5cm 1.75cm 1cm 1cm},scale=0.6]{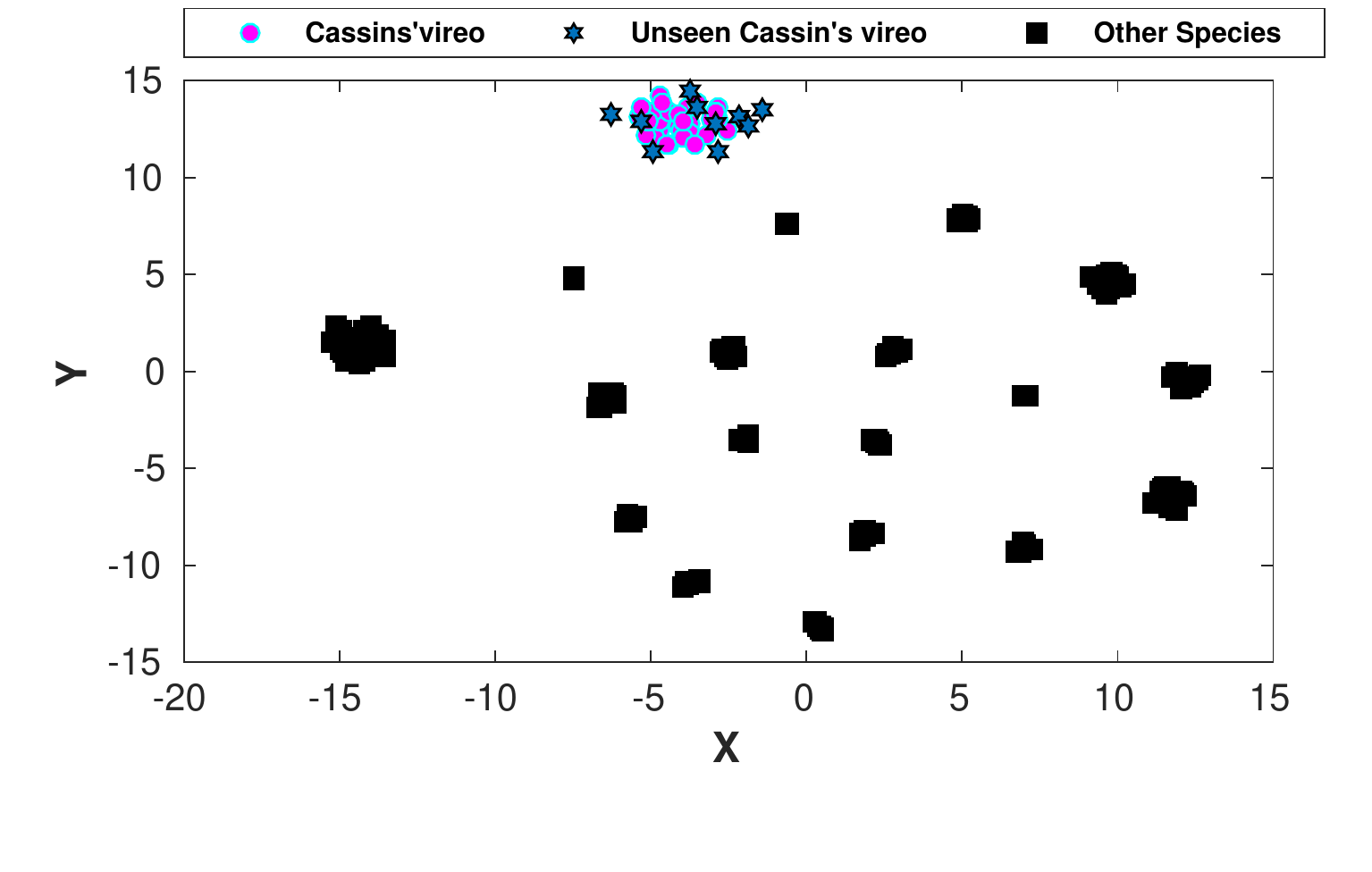}
	\caption{t-SNE visualization of embeddings generated from seen and unseen Cassin's vireo song phrases using MS-CNN-TL.     
	}
	\label{fig:gen}
\end{figure}

\subsection{Generalization of the proposed DML framework}

In bioacoustic classification tasks, the training examples often do not contain the whole repertoire of vocalizations that a species can produce. In field conditions, the test examples often contain vocalizations that are not used for training. Thus, an effective classification framework must be able to generalize on these unseen examples. To study the generalization ability of the proposed framework, the unseen vocalizations (not included in training train the model) of Cassin's vireo (one of the species in Birdcalls71 dataset) are used. The ten unseen song phrases are extracted from the audio recordings available at \url{https://goo.gl/x17fYf} and are provided for analysis along with Birdcalls71 dataset. The t-SNE representation of the embeddings extracted from these unseen song phrases using the trained MS-CNN-TN are shown in Fig.~\ref{fig:gen}. The analysis of this figure makes it clear that embeddings generated from the unseen song phrases of Cassin's vireo exhibit more similarity to the training Cassin's vireo examples than embeddings of other species. This is attributed to the fact that the triplet loss used in the proposed framework deals with grouping the similar vocalizations together and separating them from the dissimilar examples. Generally, vocalizations of a species are more similar to each other than the sounds produced by other species (though exceptions are always present in natural systems). As a result, the embeddings extracted from seen or unseen vocalizations of a species are bound to be grouped together. Thus, the utilization of deep metric learning (DML) helps in overcoming small variations in the nature of vocalizations as well as differences in the recording environment during training and testing.

\subsection{Effect of using harmonic and percussive components} 
\begin{figure*}[t]
	\centering
	\includegraphics[trim={3cm 1.75cm 3cm 1cm},scale=0.6]{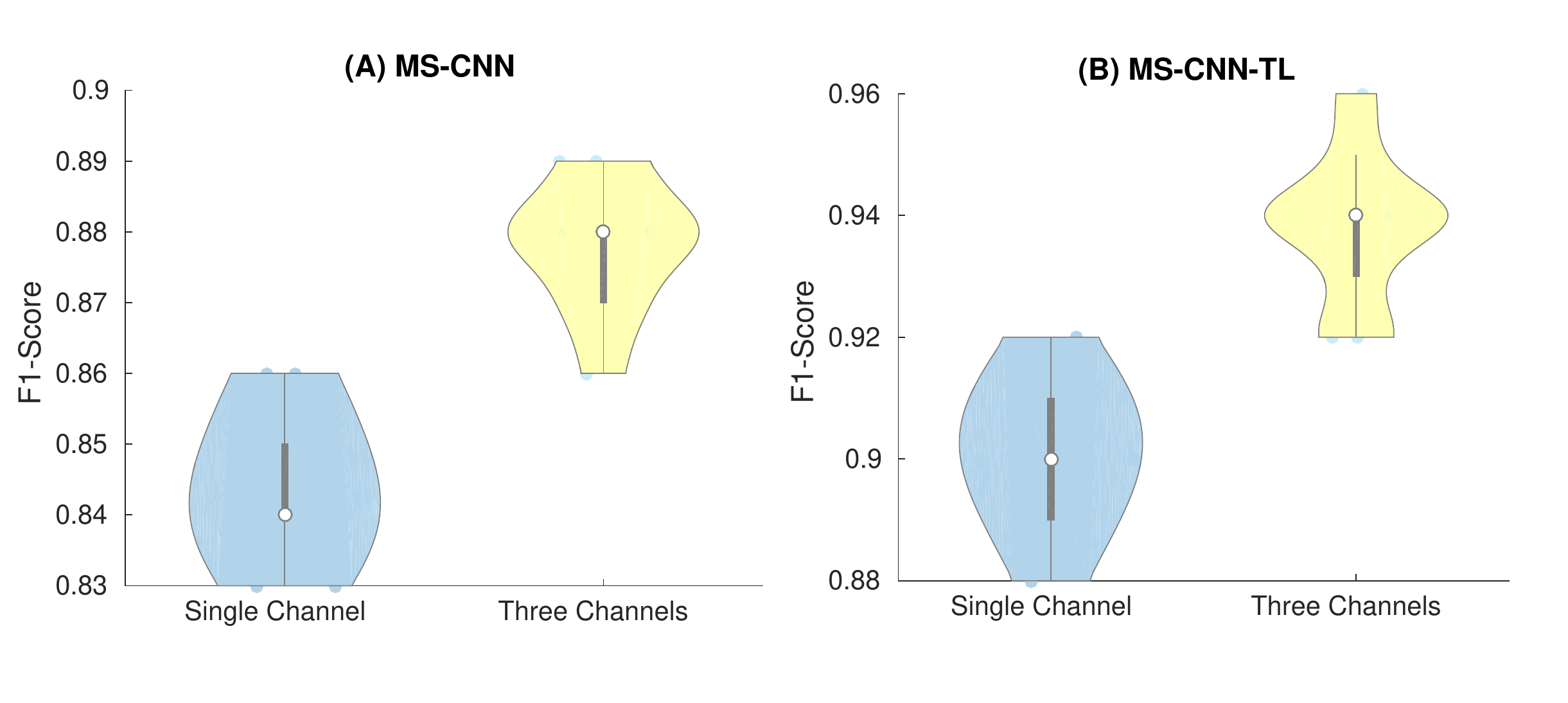}
	\caption{Violin plots depicting the effect of utilizing the harmonic and percussive components along with Mel-spectrogram on the classification performances of (A) MS-CNN and (B) MS-CNN-TL.}
	\label{fig:3_ch}
\end{figure*}

To analyze the effect of using harmonic and percussive components along with mel-spectrograms, MS-CNN (with cross-entropy loss) and MS-CNN-TL (with triplet loss) are trained on Birdscalls71 dataset using both three-channeled input and Mel-spectrograms only. In each scenario, a model is trained 10 times to compensate for the effects of random weight initialization. The violin plots depicting the classification performances achieved in both these scenarios are illustrated in Fig.~\ref{fig:3_ch}. The analysis of these violin plots makes it clear that utilizing percussive and harmonic components along with Mel-spectrograms helps in achieving a better classification performance. This demonstrates that the differences in harmonic and percussive components of animal sounds can be exploited for species classification. 



\subsection{Exploring MS-CNN-TL for open set classification}
Open set classification is a challenging issue in designing bioacoustic classification frameworks for field conditions. Current frameworks including the proposed MS-CNN-TL assign any input test example to a class that exhibits maximum similarity, even if this test example does not belong to any of the classes involved in training. Thus, there must be a way to reject such examples without affecting the classification performance. To tackle open set classification, we propose to use the metric learning module of MS-CNN-TL. First an input test example is fed to MS-CNN-TL and an output label is obtained. Now, if the embedding of test example lies at a significant distance from training embeddings of the predicted class, it can be considered as an outlier and must be rejected.

To model the distance from training embeddings, an unimodel Gaussian distribution is utilized. The embeddings of training and validation examples of each class are obtained from the trained MS-CNN-TL. The training embeddings are averaged to obtain a mean vector. A Gaussian distribution is fitted over the distance of validation embeddings from this mean vector. The maximum likelihood estimation is used to calculate the mean and variance of these Gaussian distributions. During testing, first an input test example is classified by MS-CNN-TL. Then, the distance of the test embedding from the mean vector of the predicted class is calculated. Finally, the likelihood of this distance is computed with respect to the Gaussian distribution of the predicted class. If this likelihood is less than $0.5$, the test example is considered as outlier and is rejected.

\begin{table}[h]

\caption{Classification and outlier rejection performances of the proposed MS-CNN-TL framework in different training-testing setup. In each setup, a dataset is used for training and classification evaluation whereas a different dataset is used for evaluating the outlier rejection mechanism.}
\label{tab:out}
\resizebox{0.45\textwidth}{!}{ 
\begin{tabular}{|c|c|c|c|c|}
\hline
\textbf{}                                                            & \multicolumn{2}{c|}{\textbf{\begin{tabular}[c]{@{}c@{}}Classification\\ Setup\end{tabular}}}                                                                                   & \multicolumn{2}{c|}{\textbf{\begin{tabular}[c]{@{}c@{}}Outlier Rejection\\ Setup\end{tabular}}}                                                     \\ \hline
\textbf{\begin{tabular}[c]{@{}c@{}}Training \\ Dataset\end{tabular}} & \textbf{\begin{tabular}[c]{@{}c@{}}Testing \\ Dataset\end{tabular}} & \textbf{\begin{tabular}[c]{@{}c@{}}Classification\\ Performance \\ (Macro F1-score)\end{tabular}} & \textbf{\begin{tabular}[c]{@{}c@{}}Outlier \\ Dataset\end{tabular}} & \textbf{\begin{tabular}[c]{@{}c@{}}Rejection \\ Accuracy\\ (\%)\end{tabular}} \\ \hline
Birdcalls71                                                          & Birdcalls71                                                         & 0.91                                                                                              & Anuran                                                              & 97                                                                            \\ \hline
Anuran                                                               & Anuran                                                              & 0.95                                                                                              & Birdcalls91                                                         & 93                                                                            \\ \hline
\end{tabular}}
\end{table}

A small experiment is designed to analyze the rejection accuracy of the above mentioned framework. First, MS-CNN-TL is trained on Birdcalls71 dataset and Anuran dataset is used for outlier rejection. Then, the framework is trained on Anuran dataset and Birdcalls71 is used for outlier rejection. The results of this experiment are documented in Table \ref{tab:out}. The analysis of this table makes it clear that in both setups, MS-CNN-TL with the aforementioned rejection mechanism is able to reject the outliers with good accuracy of $97\%$ and $95\%$. However, a small relative drop in macro F1-scores is observed. This shows that as expected, incorporating outlier rejection mechanism in MS-CNN-TL leads to a small drop in classification performance. However, this observed classification performance is still competitive in comparison to the other methods considered in this study (see Fig.~\ref{fig:results}).

\section{Conclusion}
\label{sec:con}
In this paper, the authors presented a deep metric learning based framework for bioacoustic classification. The authors proposed a multiscale CNN and a dynamic triplet loss to achieve effective deep metric learning even in scarcity of the training data. The proposed multiscale CNN utilizes different kernel sizes to extract features at different granularities. Whereas, the nature of dynamic triplet loss significantly increases the amount of triplets during course of training. The embeddings extracted from multiscale CNN based DML are used as a feature representation for classifying an input example. The experimental results on four different datasets show that the proposed DML based classification framework performs better than existing bioacoustic classification frameworks and various CNN architectures trained using cross-entropy loss. The authors also presented a simple augmentation that enables the proposed framework to perform open-set classification. 

A major drawback of the proposed framework (and most of the existing metric learning frameworks) is that it can not handle multi-label classification. Future work may involve developing metric learning frameworks to overcome this drawback.

\begin{acknowledgments}
This work is partially supported by IIT Mandi under the
project IITM/SG/PR/39 and Science and Engineering
Research Board, Government of India under the project
SERB/F/7229/2016-2017

The authors thank Dr~Vincent Lostanlen for helpful discussions on data augmentation in context of bioacoustics and bird vocalizations.
\end{acknowledgments}

\end{document}